\documentclass[review]{elsarticle}
\pdfoutput=1
\usepackage{lineno,hyperref}

\begin{document}

\begin{frontmatter}

\title{Properties of GaAs:Cr-based Timepix detectors} 


\author[mymainaddress,mysecondaryaddress]{P. Smolyanskiy\fnref{myfootnote}}
\ead{smolyanskiy@jinr.ru}
\author[mymainaddress]{G. Chelkov}
\author[mymainaddress]{S. Kotov}
\author[mymainaddress]{U. Kruchonak}
\author[mymainaddress]{D. Kozhevnikov}
\author[mythirdaddress]{S. Pospisil}
\author[mythirdaddress]{I. Stekl}
\author[mymainaddress]{A. Zhemchugov}


\address[mymainaddress]{Joint Institute for Nuclear Research, Dubna, Russia}
\address[mysecondaryaddress]{Saratov State University, Saratov, Russia}
\address[mythirdaddress]{Institute of Experimental and Applied Physics in Prague, Czech Republic}

\fntext[myfootnote]{Corresponding author}

\begin{abstract}

It is the hybrid pixel detector technology which brought to the X-ray imaging a low noise level at a high spatial resolution, thanks to the single photon counting. However, silicon as the most widespread detector material is marginally sensitive to photons with energy more than 30~$keV$. That's why the high-Z alternatives to silicon such as gallium arsenide and cadmium telluride are increasingly attracting attention of the community for the development of X-ray imaging systems in recent years. We present in this work the results of our investigations of the Timepix detectors bump-bonded with sensors made of gallium arsenide compensated by chromium (GaAs:Cr). The properties which are mostly important from the practical point of view: IV~characteristics, charge transport characteristics, operational stability, homogeneity, temperature dependence as well as energy and spatial resolution are considered. Applicability of these detectors for spectroscopic X-ray imaging is discussed. 

\end{abstract}


\end{frontmatter}


\section{Introduction}

Semiconductor particle detectors work as solid state ionization chambers and their principle of operation is very similar to that of gas ionization particle detectors. The advantage of semiconductor detectors over gas ionization detectors consists in the fact that the density of the sensor material in semiconductor detectors is about three order of magnitude higher than the density of gas (e.g. the density of argon is $\rho_{Ar} = 1.8 \times 10^{-3}$ $g/cm^{3}$ whereas the density of silicon is $\rho_{Si} = 2.3$ $g/cm^{3}$). It leads to higher number of charge carriers generated per particle track length in case of using semiconductor sensors. Also the energy required for creation of charge carriers in semiconductor is almost an order of magnitude smaller than the energy required for ionization of gas molecules.

Hybrid pixel detector technology initially developed for particle tracking at CERN has been finding more applications in various fields of science. With respect to X-ray imaging this technology allows to have a low noise level at a high spatial resolution thanks to the principle of single photon counting. Also with hybrid pixel detector technology it is possible to choose sensor materials which are best suited to the requirements imposed by the application. Silicon is the most widespread semiconductor sensor material for hybrid pixel detectors since it is well studied, relatively low in cost and available in large sizes. But, silicon is not suitable for X-ray imaging applications with energies of photons more than 30~$keV$ because of its low atomic number. So, heavier alternatives to silicon such as gallium arsenide and cadmium telluride have increasingly being used as sensor material for hybrid pixel detectors in such X-ray imaging systems with photon energies up to 100~$keV$. The basic properties of semiconductor materials are presented in Table~\ref{table1}). Germanium can be used for such applications too but it has to be cooled due to the high leakage current at room temperatures. This leads to more complicated and expensive construction of the imaging systems. 

\begin{table}[ht]
\caption{\label{table1} Basic properties of semiconductor materials}
\begin{center}
  \begin{tabular}{|c|c|c|c|} \hline
    Material & Density [$g/cm^3$] & Atomic number & Band gap [$eV$] \\ \hline
    $Si$ & 2.33 & 14 & 1.12 \\   
    $GaAs$ & 5.32 & $\sim$32 & 1.43 \\ 
    $Ge$ & 5.32 & 32 & 0.66 \\ 
    $CdTe$ & 5.85 & $\sim$49 & 1.50 \\ \hline
\end{tabular}
\end{center}
\end{table}

\section{Gallium arsenide as detector material}

Gallium arsenide has long seemed a promising material~\cite{doḡru1994lec}, but despite numerous attempts, it was impossible to develop good quality detectors of ionizing radiation until recently. In the world practice wafers of semi-insulating gallium arsenide produced by means of the Liquid Encapsulated Czochralski method (LEC~SI-GaAs) are usually used. Nevertheless, LEC~SI-GaAs has significant drawbacks which limit its use as a material for particle detectors. The main problems are a high leakage current and a nonuniform distribution~\cite{tyazhev2003gaas} of the electric field. However, the technology for compensation of n-type conductivity in LEC~SI-GaAs layers by thermal diffusion of deep acceptor impurity of chromium atoms (GaAs:Cr) was developed by a scientific group of the Tomsk State University (TSU)~\cite{tyazhev2003gaas}. The experimental values of the resistivity for GaAs:Cr structures are about $10^9$ $\Omega\cdot cm$~\cite{tyazhev2003gaas,ayzenshtat2004gaas}, which is more than an order of magnitude greater than the resistivity of structures based on LEC~SI-GaAs~\cite{tyazhev2003gaas}. Due to the high resistivity of the material there is no need for Schottky contacts and thus ohmic contacts is employed leading to the linear IV~characteristics. Furthermore, it was shown in~\cite{tyazhev2003gaas} that the electric field is uniformly distributed through the entire thickness of the sensor. To the date, the TSU group is able to produce 4-inch GaAs:Cr wafers with the thickness of sensitive layer up to 1~$mm$. 

\begin{figure}[htbp]
\centering 
\includegraphics[width=.6\textwidth]{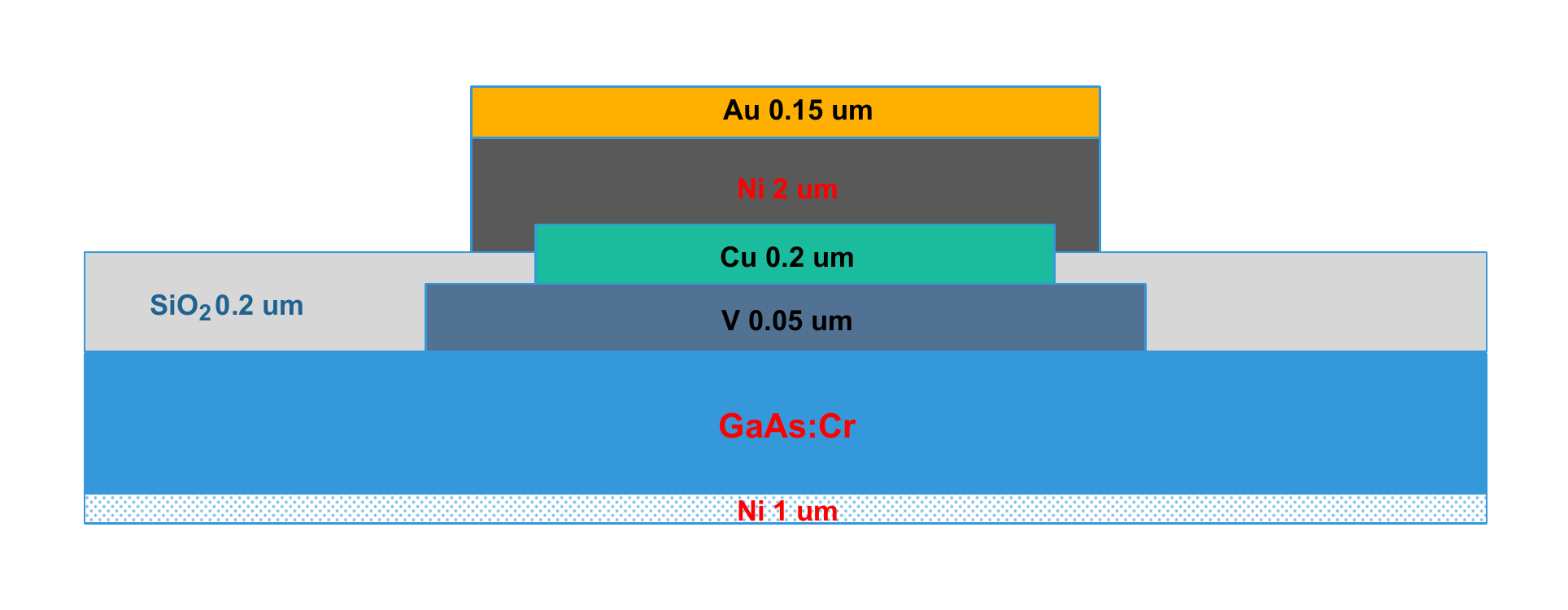}
\qquad
\includegraphics[width=.3\textwidth]{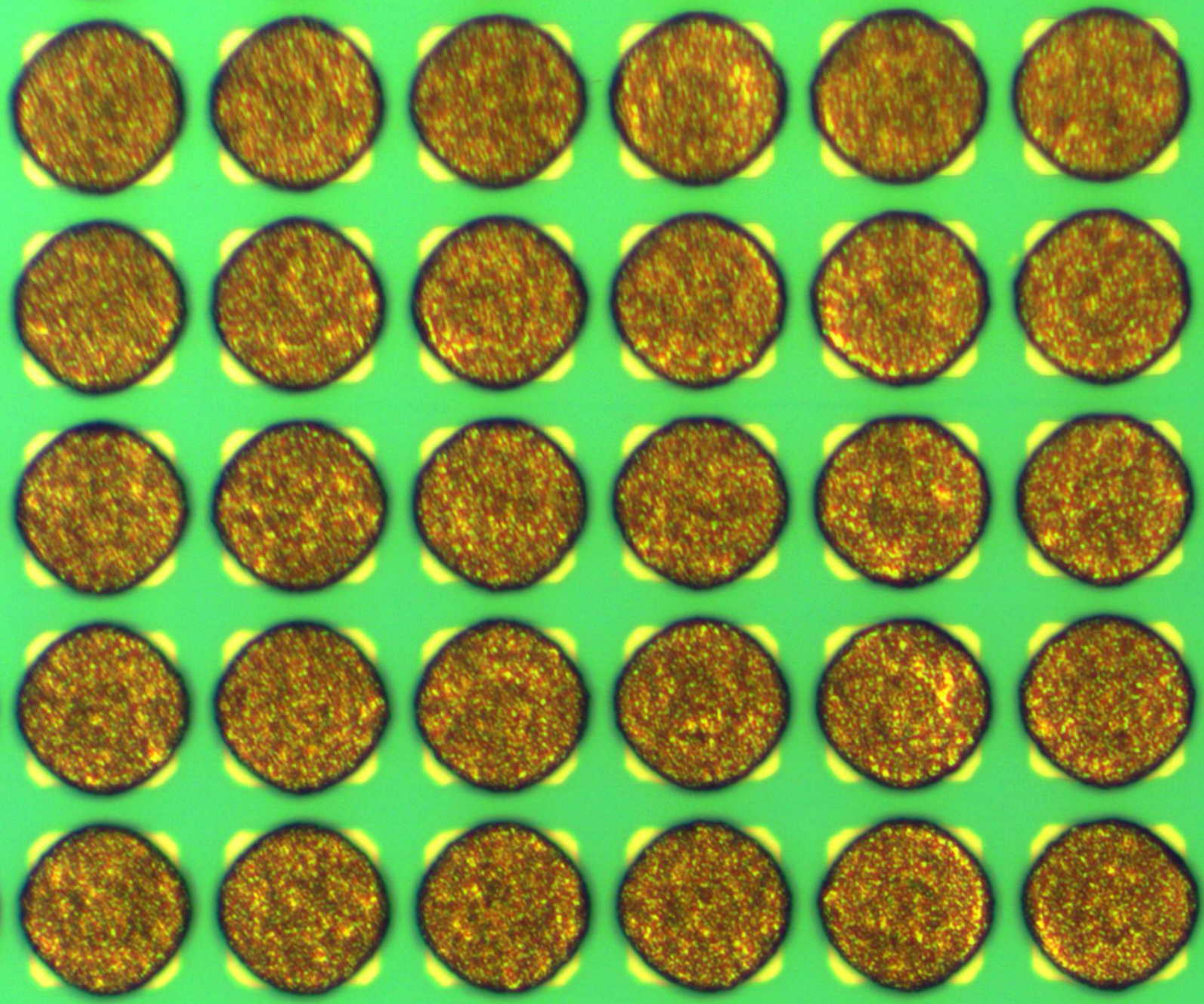}
\caption{\label{fig_pixels} (left) Structure of pixel and common contacts at GaAs:Cr sensors. (right) Solder bumps placed on the sensor UBM.}
\end{figure}

The structure of anode-GaAs:Cr-cathode contacts made for detectors using in this work are shown in Figure~\ref{fig_pixels} (left). Common contact made from 1~$\mu m$ thick Ni layer, whereas pixel contact side is produced from layers of V, Cu, Ni, Au to provide better adhesion to the GaAs:Cr material. To some sensors the guard rings with width of 150~$\mu m$ are applied to reduce leakage current on the sensor edges. Solder bumps placed on the sensor before flip-chip bonding are depicted in Figure~\ref{fig_pixels} (right).

The idea of applying the GaAs:Cr material in production of hybrid pixel detectors belongs to the JINR group.  The first prototype of the GaAs:Cr-based Medipix2 detector was fabricated in 2008 and despite being the first attempt it has demonstrated very promising results~\cite{tlustos2011characterisation}. Henceforth the JINR group together with the Tomsk State University, Medipix collaboration, Institute of Experimental and Applied Physics in Prague is working on developing and testing hybrid pixel detectors based on the Timepix readout chip and sensor matrix of GaAs:Cr. Main characteristics of such detectors were measured and the results are reported in this article. The article is a kind of summary of GaAs:Cr-based detectors investigations performed by JINR group together with colleagues over the last five years. Moreover, the emergence of a new generation of Medipix family chips allowed to expand the results shown in~\cite{tlustos2011characterisation}. Some of our results presented in the article are already published but for completeness they are given here briefly with the latest updates. 

\section{Timepix detector}

All detectors characterized in this work consist of a GaAs:Cr sensor (with a thickness of 300/500/1000~$\mu m$) bump bonded to the Timepix readout chip~\cite{llopart2007timepix}. It has a matrix of 256x256 square pixels with a pitch of 55~$\mu m$, which gives the physical size of the sensitive area of 14.1x14.1~$mm^2$. Each pixel of the chip features a charge sensitive preamplifier, a threshold discriminator and a 14-bit pseudo-random counter. The Timepix chip is compatible with both polarities of input signal and through the Krummenacher~\cite{krummenacher1991pixel} architecture of the front-end electronics is capable to compensate leakage currents from the sensor material of up to 10~$nA/pixel$ or 20~$nA/pixel$ in $e^{-}$ or $h^{+}$ collection modes, respectively. 

Depending on chip operation mode the output preamplifier signal of each pixel can be processed in three different ways:
\begin{itemize}
\item in counting mode (or Medipix mode) the chip counts how many times during the open shutter the signal exceeds a certain threshold i.e. the number of incoming particles;
\item in Time-over-Threshold (TOT) mode the chip measures for how long the signal stays above the certain threshold\footnote{The time over threshold is measured by well-defined Timepix clock in terms of TOT values}, and thanks to the pulse shape this time period can be translated to energy via appropriate calibration~\cite{jakubek2011precise,butler2015measurement};
\item in Time-of-Arrival (TOA) mode the chip measures time from the moment when the signal crosses the threshold until the shutter is closed;
\end{itemize}

For the detector control and data taken performed in this work the readout interfaces Relaxd~\cite{visser2011gigabit} and FitPix~\cite{kraus2011fitpix} were used with corresponding SoPhy and Pixet Pro control softwares.

\section{Measurements of IV characteristics}

The IV~characteristics, as one of the basic parameters of a semiconductor device, were measured with source meter Keithley2410 at different temperatures for a 300~$\mu m$ thick GaAs:Cr sensor already assembled with a Timepix readout chip\footnote{The chip was switched off}. The detector was placed in a thermostat system that was capable of maintaining the set temperature in the range from -10~$^{o}C$ to 70~$^{o}C$ with sufficient stability ($<$ 0.2~$^{o}C$).

\begin{figure}[htbp]
\centering 
\includegraphics[width=.45\textwidth]{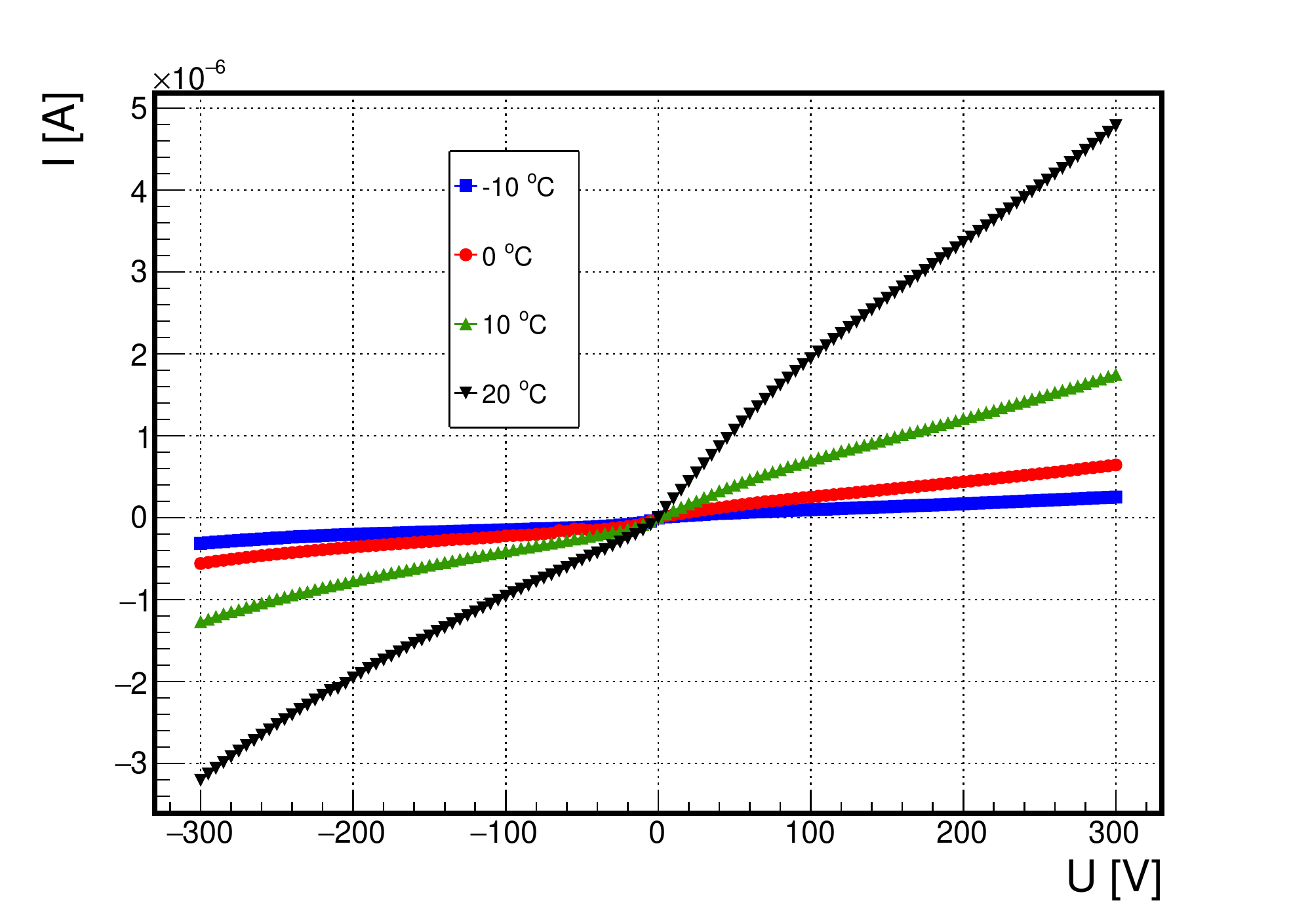}
\qquad
\includegraphics[width=.45\textwidth]{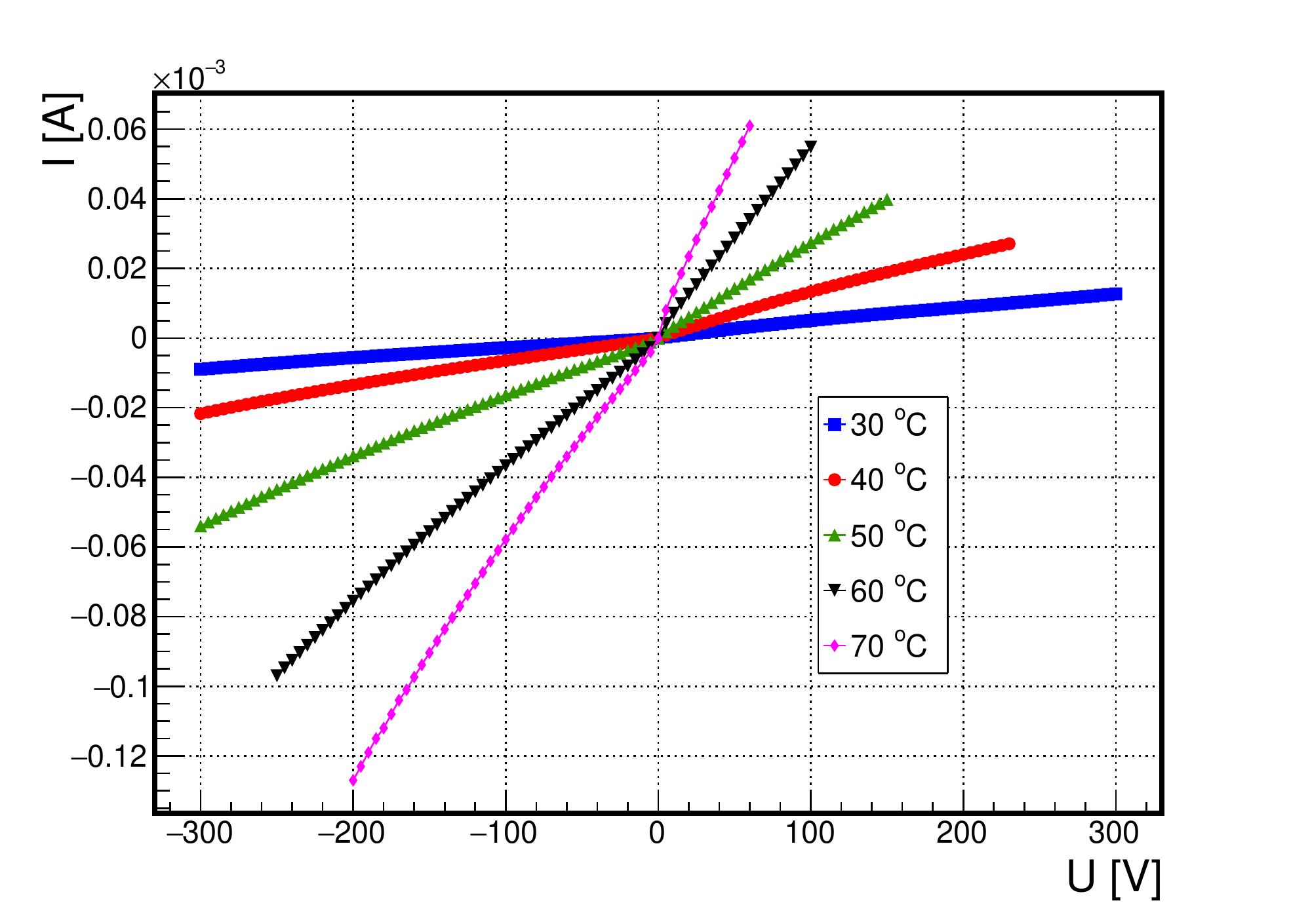}
\caption{\label{fig_IV_diffTemp} (left) IV curves for temperature range [-10, 20] $^{o}C$. (right) IV curves for temperature range [30, 70] $^{o}C$. All these IV curves are measured for the GaAs:Cr detector with dimensions of 14.1x14.1x0.3 $mm^3$.}
\end{figure}

In Figure~\ref{fig_IV_diffTemp} the volt-ampere characteristics of the detector are depicted. They are almost linear for the both polarities of the bias voltage and for all temperatures that proves the resistive behavior of the semi-insulating material. At the temperature of 20~$^{o}C$ and the bias voltage of -300~$V$ the detector has leakage current of 3.2~$\mu A$. If we assume that the total current is evenly distributed among all pixels the leakage current per pixel is about 0.05~$nA$. The Timepix front-end electronics can compensate with a margin such leakage current, therefore chip can operate with GaAs:Cr sensors normally.

\begin{figure}[htbp]
\centering
\includegraphics[width=.8\textwidth]{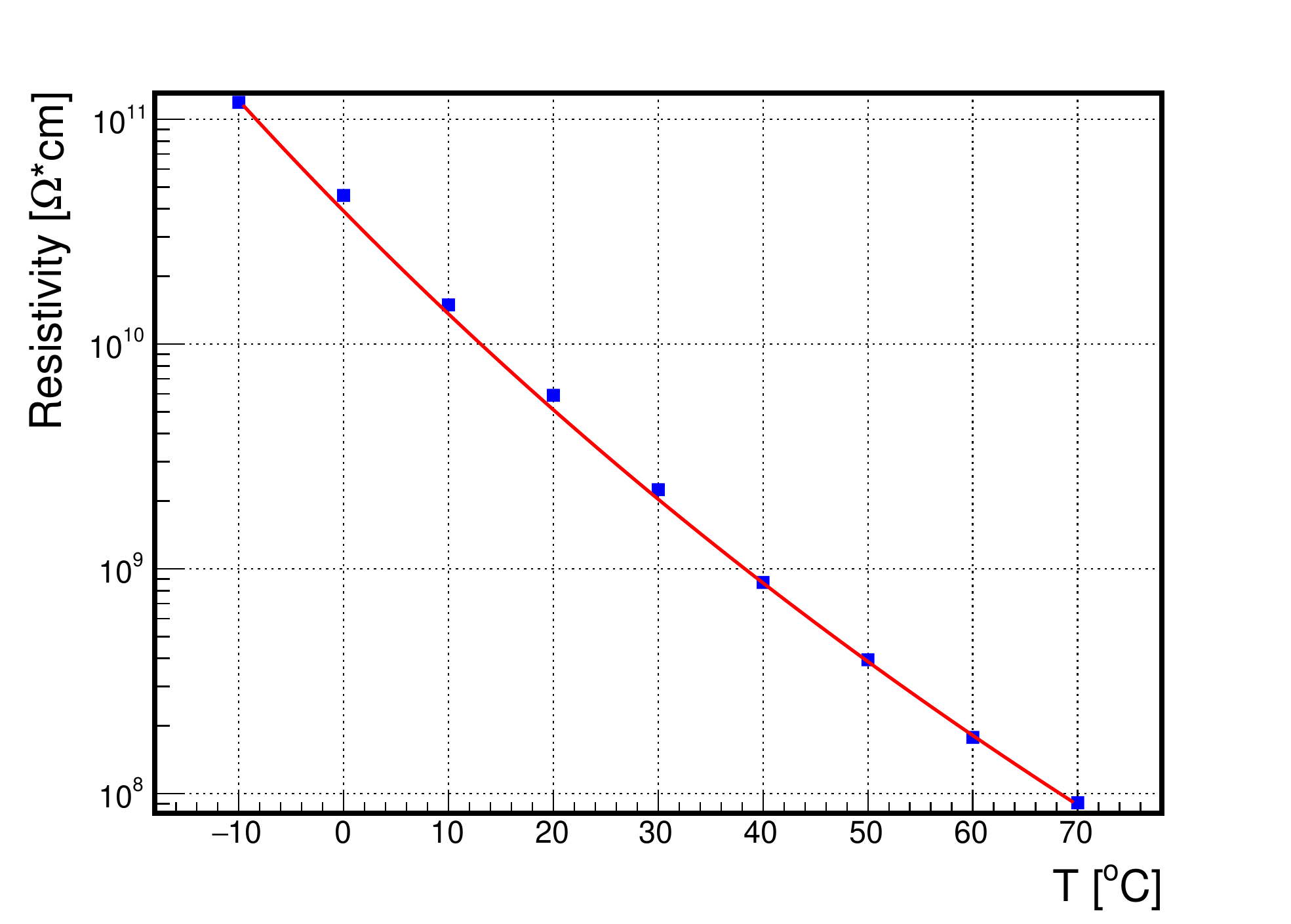}
\caption{\label{fig_R_T} Temperature dependence of the GaAs:Cr material resistivity for negative polarity of the bias voltage.}
\end{figure}

From the measured IV characteristics the electrical resistivity dependence on the temperature was calculated. Such dependence for negative branch of the bias voltage is shown in Figure~\ref{fig_R_T}. At the temperature of 20~$^{o}C$ the material has the electrical resistivity of about $10^8\ \Omega \cdot cm$.

For semiconductors the resistivity $\rho$ is given by analytical expression:
$$ \rho \sim e^{E_{0} / 2kT} $$
where $E_{0}$ -- the energy gap between the valence and conduction bands, $T$~--~temperature, and $k$ -- the Boltzmann constant. Using this expression it is possible to calculate the GaAs:Cr band gap value as the slope coefficient of $\rho(T)$ curves. Performing the fit, the value $E_{0} = 1.43 \pm 0.012$ $eV$ is obtained that is very close to the tabulated value of 1.43 $eV$ for the semi-insulating GaAs material, but as noted by other authors~\cite{veale2014chromium} the value of $E_{0}$ varies from sensor to sensor.

\section{Electron charge transport}

Majority charge carriers in the GaAs:Cr are electrons due to very low lifetime of holes (according to~\cite{ayzenshtat2002charge} it is 0.2~$ns$). Widely known method for the $\mu \tau$-product evaluation in pixel detectors is the irradiation of the common electrode of a detector with $\alpha$ particles and measurement of the signal generated by induced charges on the electrode, depending on the bias voltage. Then by fitting this dependence by the Hecht equation~\cite{hecht1932mechanismus}, one can obtain the value of $\mu \tau$-product.

It has been shown in~\cite{hamann2013characterization} that $\alpha$ particles cannot be used for measuring the electron $\mu \tau$-product of the GaAs:Cr material due to incomplete charge collection efficiency (CCE) because of a short lifetime of charge carriers (electrons) in comparison with the lifetime of plasma created by $\alpha$ particle~\cite{hamann2015performance}. To overcome this problem we selected characteristic X-ray radiation from zirconium foil ($K_{\alpha_{1}} = 15.7$~$keV$) as particle source. Photons corresponding to this line have a mean free range in GaAs less than 25~$\mu m$, therefore the majority of interactions will occur close to the cathode. Another advantage of using low energy photons is their single-pixel stamp primary formed in the result of interaction, therefore one can measure per-pixel map of the electron $\mu \tau$-product. 

The mean value of 15.7~keV photo peak was calculated for each pixel with varying the bias voltage (only single-pixel events are evaluated). In pixel detectors with 55~$\mu m$ pitch the small pixel effect~\cite{knoll2010radiation} is significant and therefore the classical Hecht equation is not applicable. According to~\cite{lachish2017electron} the modified Hecht equation can be written as follows:
$$Q(U) = Q_{0} \cdot \frac{d}{\mu \tau \cdot (U-U_{0})} \cdot \int_{0}^{d} w(x) \cdot e^{-\frac{-x \cdot d}{\mu \tau \cdot (U-U_{0})}} dx + e^{-\frac{d^2}{\mu \tau \cdot (U-U_{0})}}, $$
where $Q_{0}$ -- charge generated by single photon, $d$ -- detector thickness,  $\mu \tau$ -- product of the charge mobility on the charge lifetime, $U_{0}$ -- bias voltage offset that takes into account the bias voltage at which the induced charge is under the given detector threshold, $w(x)$ - the weighting potential dependence on the distance from detector cathode\footnote{The weighting potential was calculated numerically using the finite element software ONELAB (Open Numerical Engineering LABoratory)~\cite{smolyanskiy2017study}}.

\begin{figure}[htbp]
\centering 
\includegraphics[width=.48\textwidth]{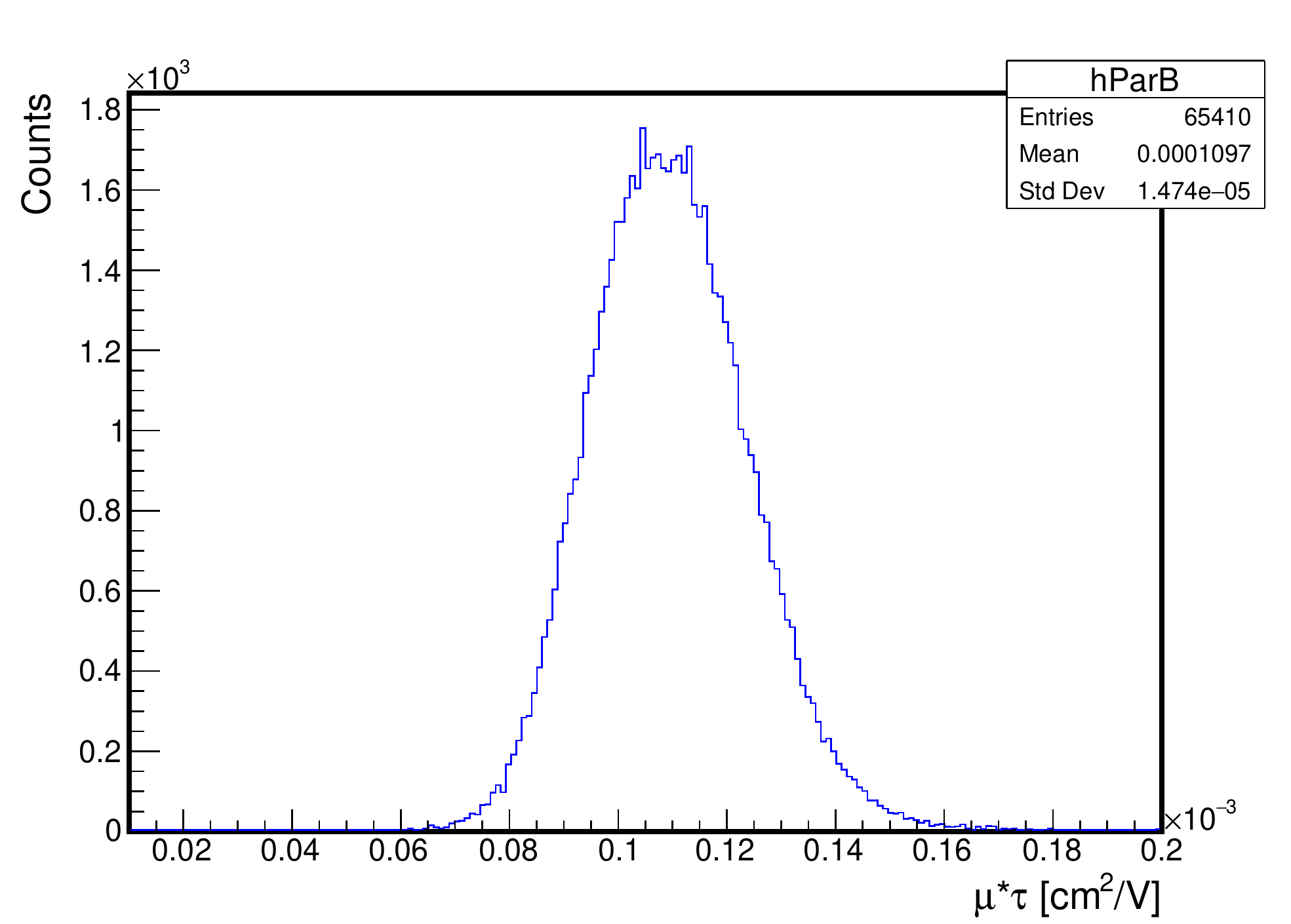}
\qquad
\includegraphics[width=.43\textwidth]{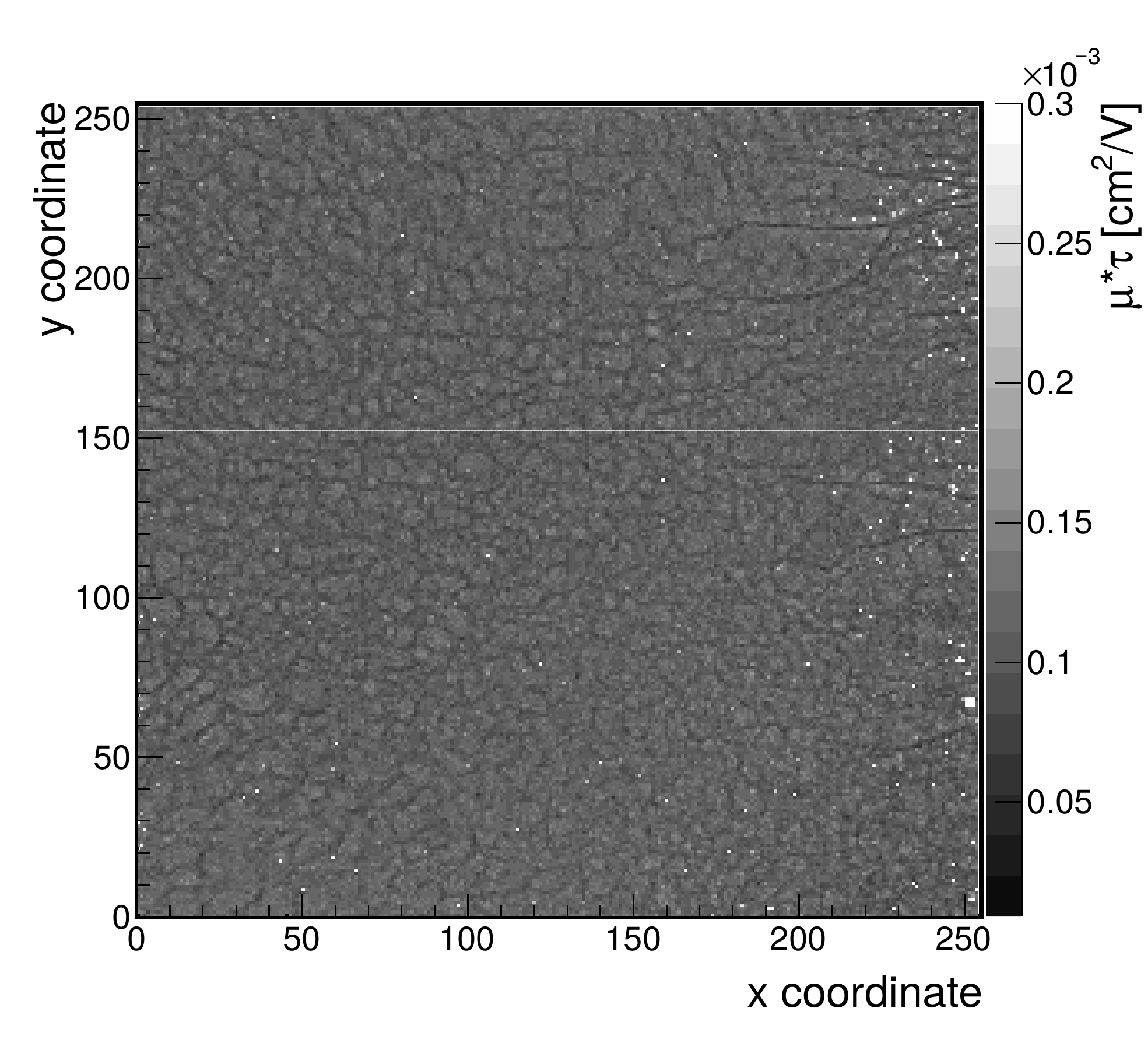}
\caption{\label{fig_MuTau} (left) Distribution of the $\mu \tau$-product for all pixels. (right) Map of the $\mu \tau$-product.}
\end{figure}

The one- and two-dimensional distributions of the $\mu \tau$-product for 500 $\mu m$ thick detector L05-W0225 are presented in Figure~\ref{fig_MuTau}. The mean value is $\mu \tau = 1.1 \cdot 10^{-4}$ $cm^{2}/V$ with $RMS = 1.47 \cdot 10^{-5}$ $cm^{2}/V$. The obtained values are within the range of values reported by other authors (e.g.~\cite{hamann2013characterization}). Also we can conclude that electron $\mu \tau$-product is quite evenly distributed across the sensor for this detector (strongly dependent on the sensor batch).

\section{Detector homogeneity}
\label{sec:homogeneity}

\begin{figure}[htbp]
\centering
\includegraphics[width=.45\textwidth]{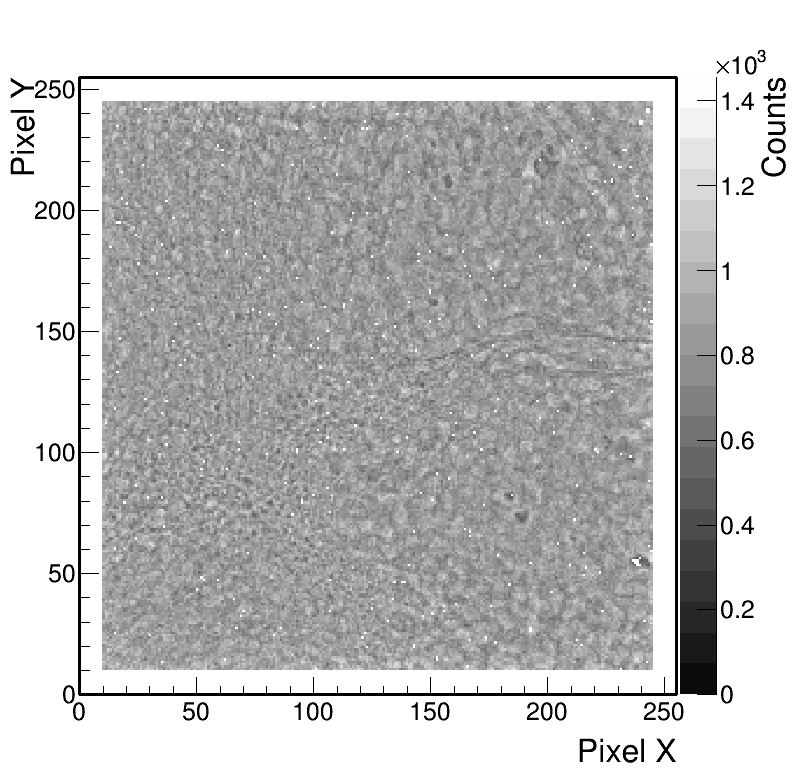}
\qquad
\includegraphics[width=.45\textwidth]{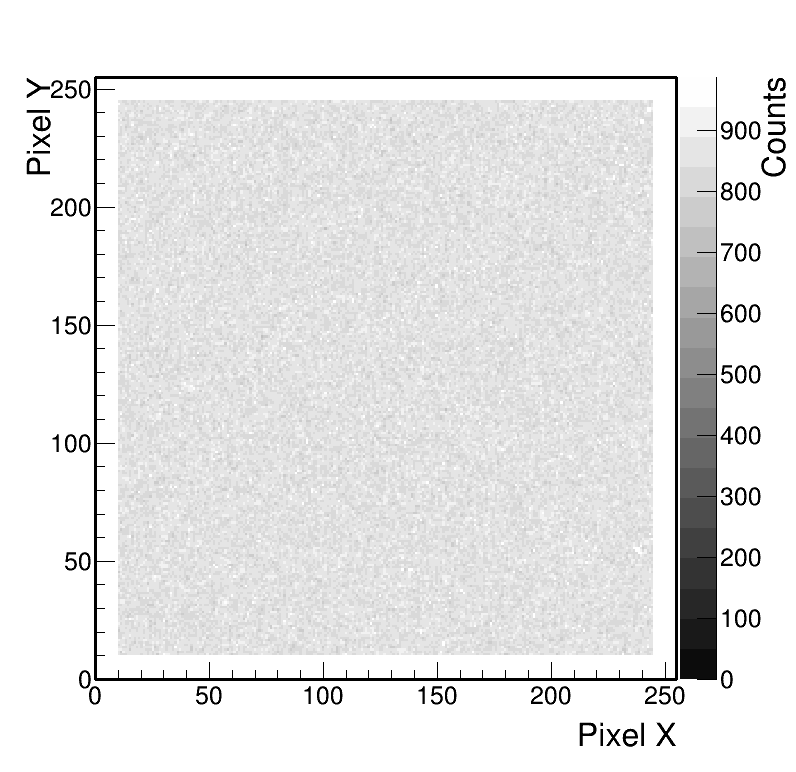}
\caption{\label{fig_ObjFrame} (left) Raw image of the GaAs:Cr-based Timepix detector. (right) Flat-field corrected image of the same assembly. Ten boundary columns and rows are masked.}
\end{figure}

\begin{figure}[htbp]
\centering
\includegraphics[width=.8\textwidth]{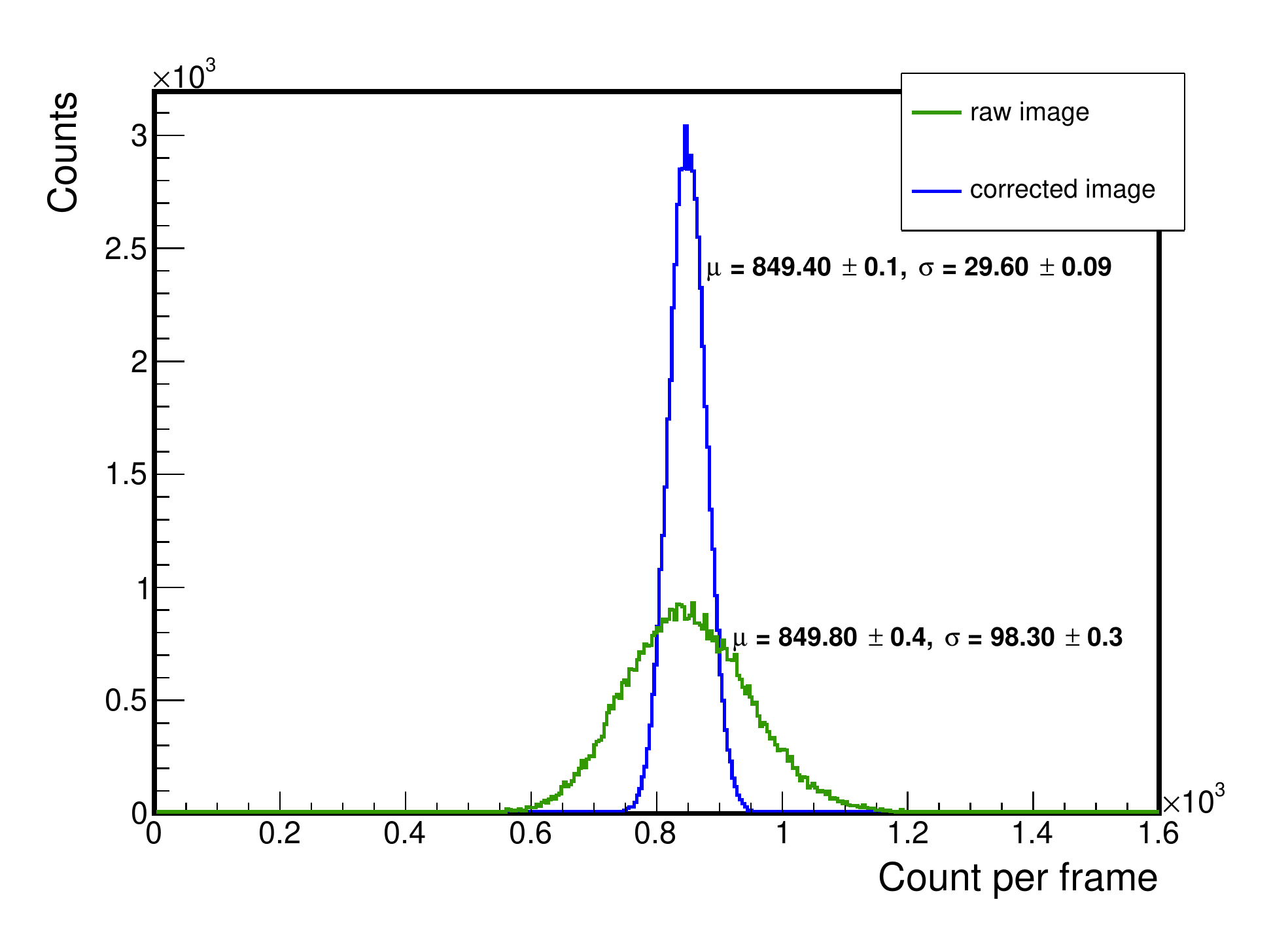}
\caption{\label{fig_CountDistr_Corr} Distribution of pixel counts over the matrix for raw and flat-field corrected images.}
\end{figure}

The bump bonding yield and the response uniformity over the detector matrix are the key characteristics which describe the quality of fabrication processes. Figure~\ref{fig_ObjFrame} (left) shows the response of a 1~mm-thick GaAs:Cr-based Timepix detector [K09-W0203]\footnote{Timepix unique chip identifier} to a uniform illumination by X-ray tube operated at 80~$kVp$. Structures corresponding to the dislocation networks of the melt-grown material are clearly visible on the image. But, these structures are stable in time and can be removed by the flat-field correction (Figure~\ref{fig_ObjFrame} (right)). This effect is shown numerically in Figure~\ref{fig_CountDistr_Corr}: distribution of pixel counts in the flat-field corrected image is more than three times narrower than in the raw image.  

The bump bonding yield of the considered assembly is about 99~\% which is much better than reported in~\cite{tlustos2011characterisation}. The statistics on fabrication yield of the GaAs:Cr detectors produced for JINR in 2015-2017 are presented in Table~\ref{table2}. The best bump bonding yield from more than 60 assemblies produced for JINR during last 5 years is hundred inactive pixels. But, unfortunately, the most of 1~$mm$ thick GaAs:Cr assemblies produced for JINR have more "bright" edges, that require to mask several boundary rows and columns\footnote{500 $\mu m$ thick assemblies have none of such issues}. This problem of close to the edge pixel lines can be connected with sensor cutting from wafer that leads to microcracks and mechanical dislocations and, ultimately, to higher leakage current.

\begin{table}[ht]
\caption{\label{table2} The GaAs:Cr detectors fabrication yield}
\begin{center}
  \begin{tabular}{|c|c|} \hline
    Purchased & 42 \\ \hline
    Rejected before bonding & 10 \\
    Assembled & 32 \\
    Bump bonding defects & 5 \\
    Sensor material defects & 3 \\ \hline
\end{tabular}
\end{center}
\end{table}

\section{Detector stability}

The degradation in performance of a sensor material or a readout chip under the action of radiation fields can lead to changes in the position of spectral peak, increase in the leakage current, etc. As one of the steps in studying a new sensor material it was important to investigate the long-term stability of spectroscopic properties of the Timepix detector with a GaAs:Cr sensor.

The long-term stability tests were done for 1~$mm$ thick GaAs:Cr-based Timepix detector [JO3-W0247] operating in the Time-over-Threshold mode. The detector placed in a lead safety housing without the light access was irradiated with a $^{241}$Am gamma source with an activity of 110 $kBq$ for the period of about one year. During the whole period of measurements the temperature was maintained at 20$\pm$0.1 $^{o}C$.

\begin{figure}[htbp]
\centering
\includegraphics[width=.8\textwidth]{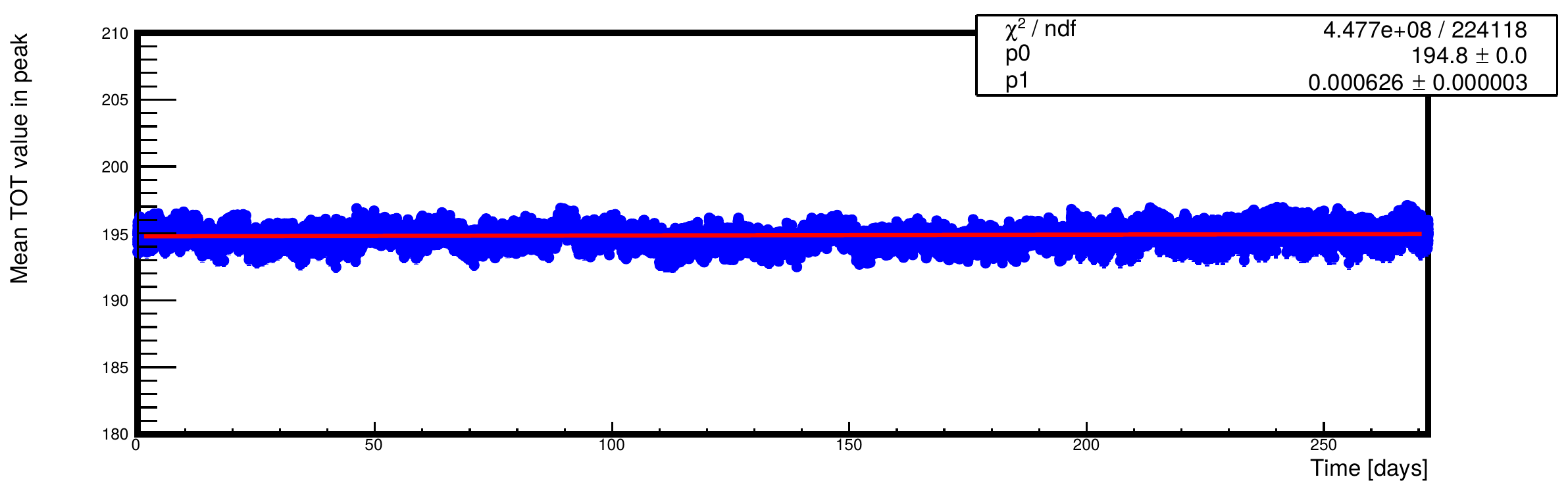}
\caption{\label{fig_Longterm} Mean peak position of the 59.5~keV gamma peak versus time. Linear fit is shown in red color.}
\end{figure}

500 frames with an interval of ten minutes were taken into account to fill the generalized energy spectrum of the $^{241}$Am source in terms of TOT counts. Afterwards, the 59.5~keV gamma peak was fitted with a gaussian for each spectrum. As a result, the mean peak position dependence on time was plotted that is shown in Figure~\ref{fig_Longterm}.

Experimental points are fitted by linear function. The mean peak position value (194.8~[TOT] with the standard deviation for whole period of 0.7~[TOT]) has increased by less than one promile per year. Thirty pixels have become inactive and several pixels have become noisy after about one year of everyday irradiation.  

Stable performance of the detectors over long time under irradiation by photon fluxes of different intensity is the key property for any X-ray imaging system. The count rate stability in the Medipix mode of a 1~$mm$ thick GaAs:Cr-based Timepix detector [K09-W0203] was investigated at two time scales for X-ray tube fluxes typical for $\mu$-CT applications.

The detector was placed in a X-ray safety cabinet at the distance of 30~$cm$ from X-ray tube\footnote{X-ray tube SB-120-350 produced by Source-Ray inc. was used for this type of measurements}. The X-ray tube was operating with voltage of 100~$kVp$ and anode current was in range of [100, 350] $\mu A$. Temperature conditions in the cabinet were constant at the level of 24~$^{o}C$. The best proof of the temperature stability is the detector current variations, that was monitored by the source meter Keithley 2410 (mean total detector current was measured at $I = 6.4 \pm 0.05$~$\mu A$ during the operation). No special measures to cool down the detector were taken, except the fan was placed behind the chipboard to avoid overheating. 

One hour long measurements with uninterrupted data acquisition under X-ray tube irradiation with current of 100 and 350~$\mu A$ were performed. One measurement was done in mode wher frames (exposure time of 200 $ms$) were recorded with 1 $s$ long pauses between acquisitions. Si-based Timepix detector was installed near the investigated GaAs:Cr detector as the X-ray tube monitor and to compare their count rate stability.

\begin{figure}[htbp]
\centering
\includegraphics[width=.88\textwidth]{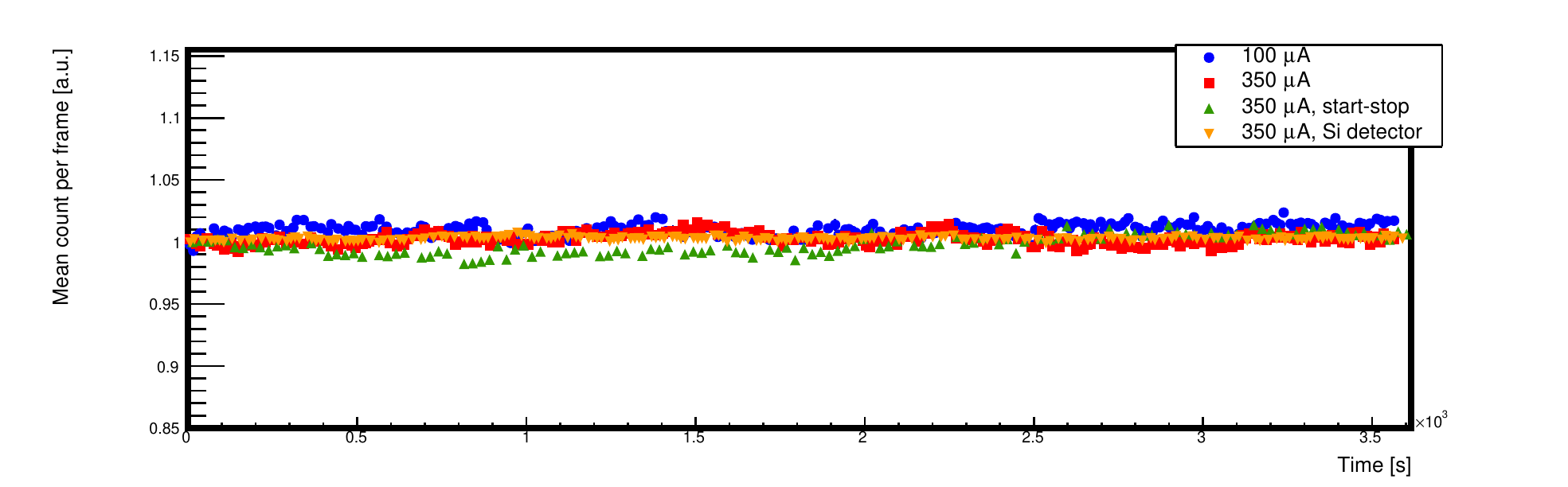}
\caption{\label{fig_MeanCountPerTime} Dependence of normalized mean pixel counts per frame on time for various X-ray tube currents and detector acquisition conditions.}
\end{figure}

Mean counts of pixels per frame were calculated, normalized to the average values from ten initial frames, and plotted versus time in Figure~\ref{fig_MeanCountPerTime}. To exclude effects on borders of the sensor, ten pixel columns and rows at the edges were withdrawn from the analysis. No significant changes in the count rates were found for the two tube currents. The RMS of mean pixels counts over time is 0.5~\% for the both tube currents. The detector demonstrates less stable behavior in mode with 1~$s$ pauses in acquisitions, but still the RMS of the mean pixel counts in time is 0.9~\%. Such deviations may be caused by cooling down of the chip during closed frame and variations of the DACs reference voltage at the opening the shutter. Si-based Timepix detector shows better count rate stability over time scale (RMS is 0.1~\%), that confirms sufficient X-ray tube stability.   

\begin{figure}[htbp]
\centering
\includegraphics[width=.88\textwidth]{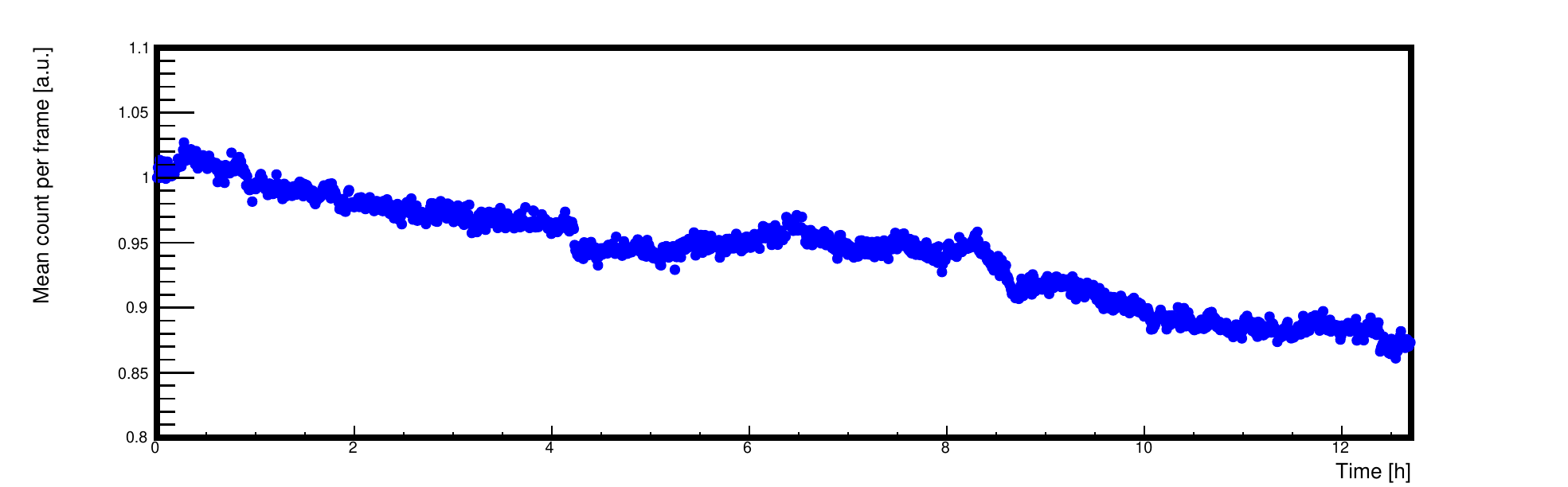}
\caption{\label{fig_MeanCountPerTime_12h} Time dependence of normalized mean counts per frame on a 12 hour scale.}
\end{figure}

The similar measurements were performed for the period of 12 hours. The X-ray tube was operating with the current of 350~$\mu A$ at voltage of 100~$kVp$ and detector has accumulated the data without pauses. In such experiment conditions the GaAs:Cr-based detector demonstrates much less stable behavior (Figure~\ref{fig_MeanCountPerTime_12h}). The mean count rate decreased by about 13~\% during the measurement. In the time period from 4 to 6.5 hours one can observe the increase of count by 2.5~\% and subsequent decrease. This count rate growth makes one think that it is not connected with instabilities of the sensor material properties, since the standard deviation still was stable within 1~\% for the whole time, as well as the temperature in the lead cabinet. A possible reason for the observed count rate fluctuations is changing in time of the reference voltage of the discriminator and other DACs, that was found by other authors~\cite{hamann2013characterization}. 

\section{Detector temperature dependence}

As the GaAs:Cr-based detectors operate at room temperature conditions it was important to investigate the temperature dependence of the detector spectral and operational characteristics. 

For these tests the same Timepix detector as in the previous section, operating at various temperatures, was irradiated with a $^{241}$Am gamma source (with an activity of 110 $kBq$). The detector and the $^{241}$Am source were located in a dark safety box. The temperature stabilization of the assembly was provided by a system based on a PID controller and a Peltier element. Such system ensured the temperature variation not more than 0.1~$^{0}С$. The tests were done in the Time-over-Threshold mode for five stable temperatures: from 20~$^{0}C$ to 25~$^{0}C$ with the step of 1~$^{0}C$. Also during one of the measurements the assembly temperature slowly varied from 25~$^{o}C$ to 20~$^{0}C$. Each measurement lasted about 1.6 hour.

\begin{figure}[htbp]
\centering
\includegraphics[width=.8\textwidth]{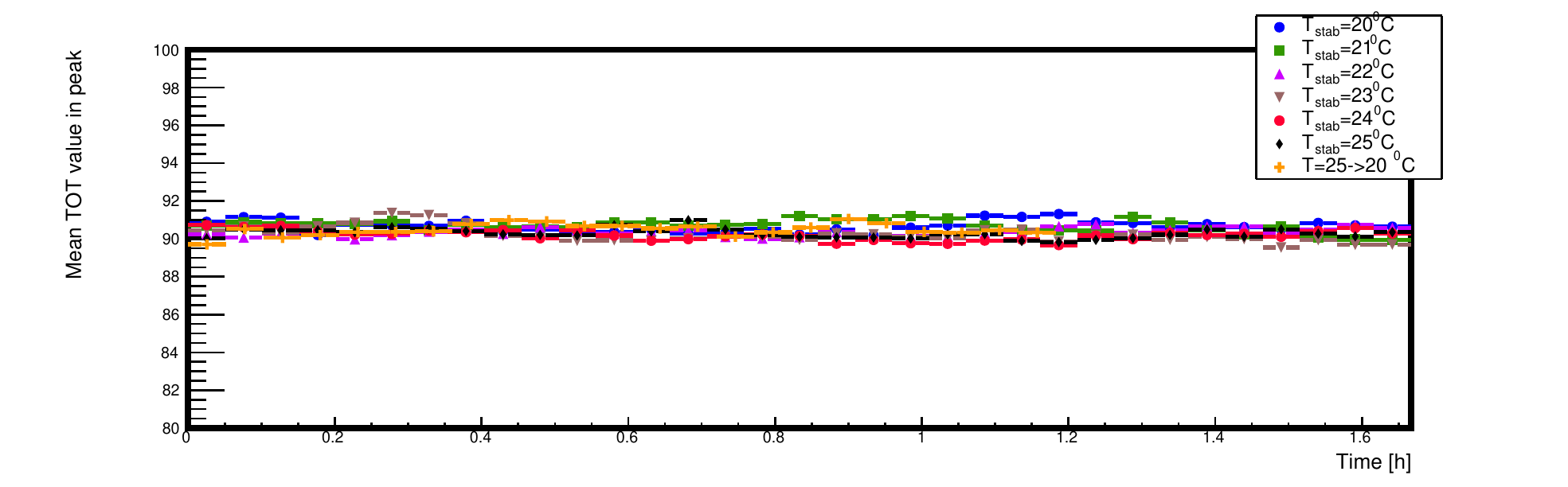}
\caption{\label{fig_PeakPosTemp} Time dependence of the detector spectral response for various  temperature conditions.}
\end{figure}

The energy spectrum of $^{241}$Am source was recorded with interval of three minute and $\gamma$ line of 59.5 keV was identified. Variations in time of spectral peak position for different temperatures are presented in Figure~\ref{fig_PeakPosTemp}. For all temperatures the detector spectral response is very stable (e.g. for 25~$^{0}C$ the mean value of the peak is 90.3 [TOT] with the RMS of 0.3 [TOT]). The spectral peak position does not change with temperature variations. Also we did not see appearance of noisy pixels with the increasing temperature, that indicates good bump bonding quality of the assembly.       

The count rate stability of the detector operating in the Medipix mode for different temperatures is at the same importance for practical applications as the detector behavior in the Time-over-Threshold mode. Similar tests were done in the Medipix mode for three stable temperatures $T_{stab}=22$~$^{0}C$, $T_{stab}=25$~$^{0}C$, $T_{stab}=28$~$^{0}C$ and, for reference, without temperature stabilization (the chip temperature was about $T=31$~$^{0}C$). 

\begin{figure}[htbp]
\centering
\includegraphics[width=.8\textwidth]{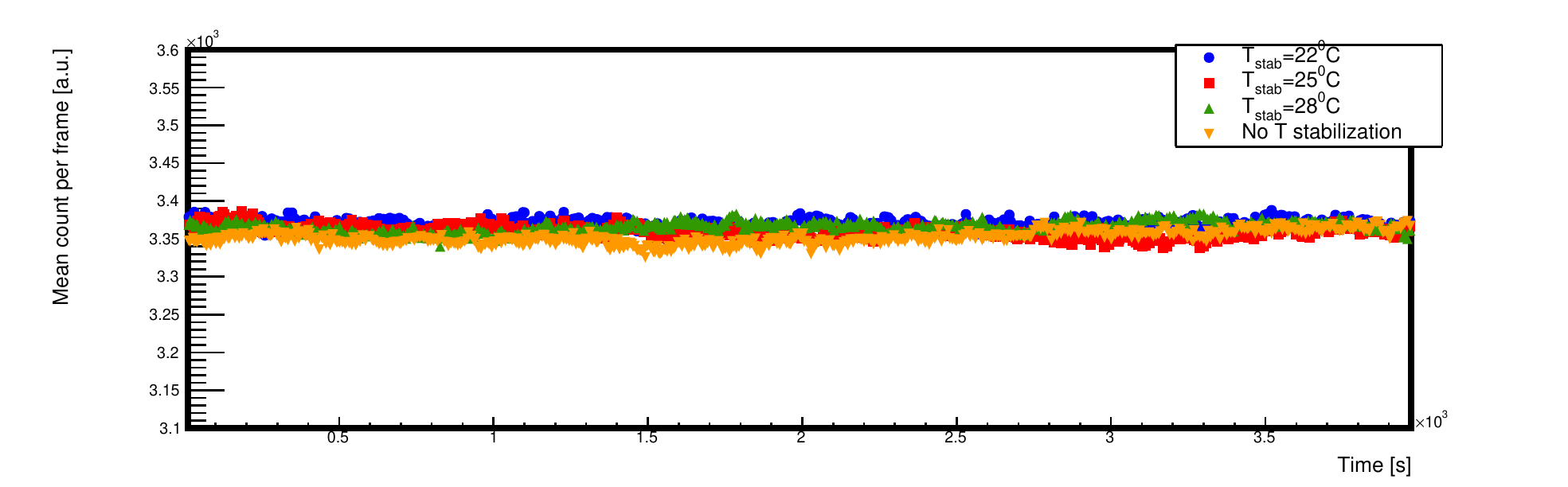}
\caption{\label{fig_MeanCountInTime_Temp} Time dependence of the detector mean counts per frame for various temperature conditions.}
\end{figure}

Figure~\ref{fig_MeanCountInTime_Temp} shows the results of such measurements. One can observe that there is no difference between count rates for three stable temperatures (standard deviation of the mean count is about 0.2~\% at one hour scale). Operation mode without temperature stabilization gives more unstable behavior but still within the standard deviation of 0.3~\%.

Based on the experience with GaAs:Cr-based pixel detectors and the results of performed tests we conclude that in most applications the temperature stabilization is not required. But, to avoid overheating of the chip it is better to put the fan behind the chipboard.  However, if temperature conditions in the experimental room is varying in a wide range (a few tens of degrees) the detector should be equipped with a temperature stabilization system.  We notice, through, that for some assemblies the cooling was necessary to cope with high noise level at room temperature due to the sensor quality.  

\section{Signal-to-noise ratio}

In X-ray photon counting detectors the number of detected photons, ideally, should follow the Poisson statistics. It means that for the mean number of absorbed photons $N$, the standard deviation of counts is defined by the square root of counts $\sqrt{N}$. To determine quantitatively the noise in the image which has the origin in the readout electronics, photon beam fluctuations, sensor inhomogeneities, the signal-to-noise ratio ($SNR$) is introduced usually. This value is defined as 
$$SNR = \frac{\mu}{\sigma},$$
where $\mu$ -- mean counts of pixels per frame, $\sigma$ -- standard deviation of pixel counts. Obviously the $SNR$ is less than $\sqrt{N}$ for real photon counting detectors.

To investigate the stability of the image quality obtained with 1~$mm$ thick GaAs:Cr-based Timepix detector [K09-W0203] several datasets were taken. The X-ray tube operated with the current of 100~$\mu A$ at the voltage of 100~$kVp$. The $SNR$ was calculated for the flat-field corrected image in dependence of number of frames taken as corrective frames. Also the frames from various time intervals of data acquisition were used as object frames for understanding the stability of the $SNR$ of the corrected image. 

\begin{figure}[htbp]
\centering
\includegraphics[width=.8\textwidth]{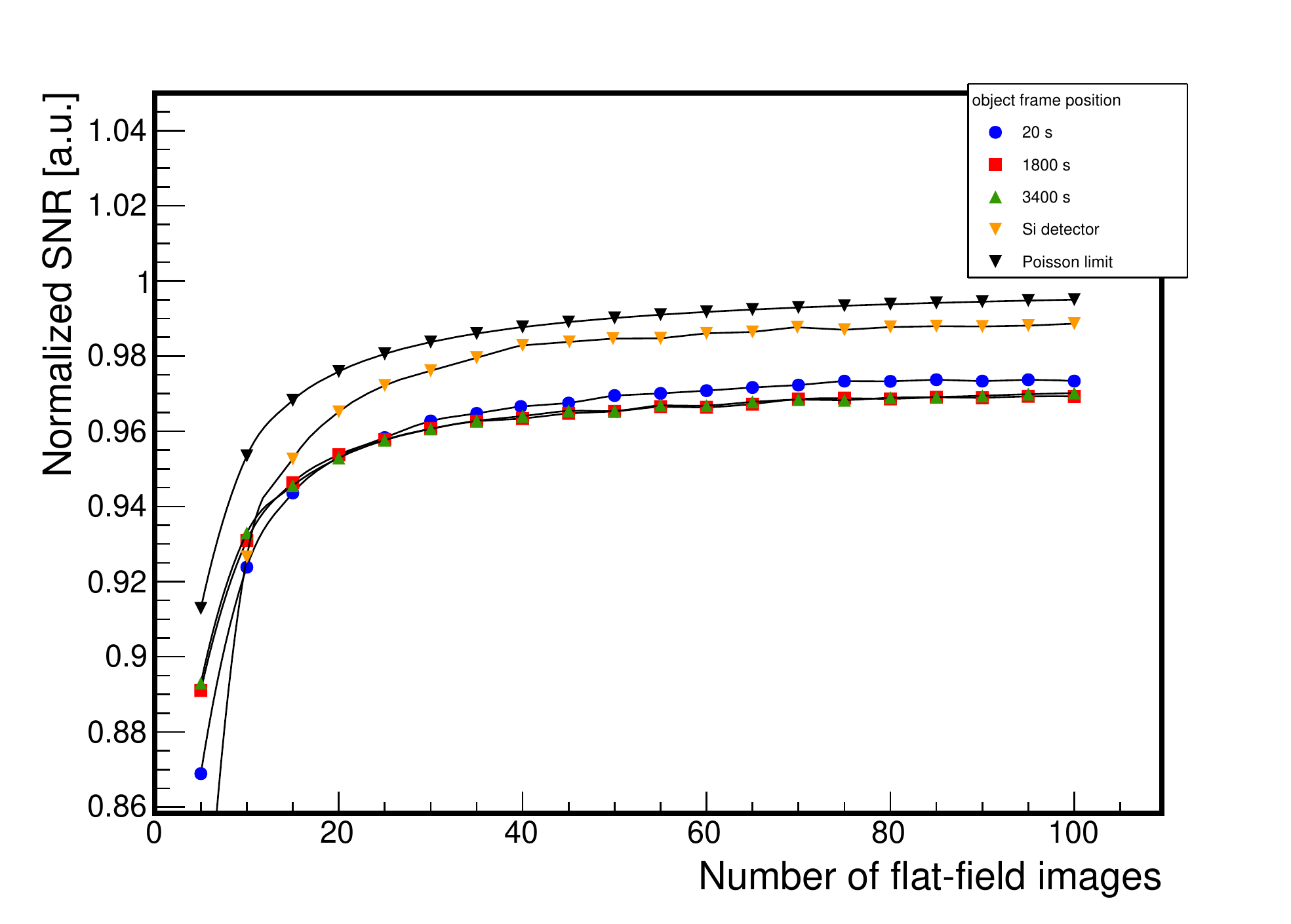}
\caption{\label{fig_SNR} $SNR$ of the flat-field corrected image in dependence of number of the corrective frames, which taken at various time intervals of data taken.}
\end{figure}

Figure~\ref{fig_SNR} shows the results of the measurements and calculations. Normalizations of curves were done to square root of the mean value of counts in the corrected frame. It shows that the Poisson limit can be approached by increasing the statistics in corrective frame. The trends of experimental curves are close to theoretical one, but experimental curves stay below the Poisson limit, which is mainly due to pixel-to-pixel threshold variations (according to~\cite{llopart2007timepix} after equalization the achieved noise free threshold variation is 35~$e^{-}$ RMS for both polarities) and residual material inhomogeneities. Such explanation is correlated with a closer to Poisson limit experimental curve obtained in the same conditions by Si-based Timepix detector (orange markers in Figure~\ref{fig_SNR}), which is well-known to be more homogeneous. The $SNR$ values calculated with corrective frames for other times presented in Figure~\ref{fig_SNR} show a very similar behavior. Thus, the obtained results confirm the conclusion made in section~\ref{sec:homogeneity} that flat-field correction is sufficient to remove almost completely the artifacts inherent in GaAs:Cr material and to achieve a high image quality. 

\section{Energy resolution}

The energy resolution of the Timepix detector operating in the Time-over-Threshold mode is determined by the following: the peaking time of the preamplifier, pixel-to-pixel variations of the gain, the clock frequency that measures the time over threshold and the sensor properties. The charge sharing contribution can be suppressed partially (signals under the threshold are not counted) by per-cluster analysis with summing of counts in neighboring pixels. But, in the Medipix mode such approach is practically not applicable due to high frame occupancy and, although this mode can give more accurate results by definition, it is difficult to determine energy resolution of thick GaAs:Cr detectors in this way.    

The energy resolution of the GaAs:Cr-based Timepix detector in the Time-over-Threshold mode after performing the per-pixel energy calibration was measured and discussed in details in~\cite{butler2015measurement}, but the recent progress in hybridization process and sensor uniformity allow us to achieve better energy resolution than presented in that paper.

\begin{figure}[htbp]
\centering
\includegraphics[width=.8\textwidth]{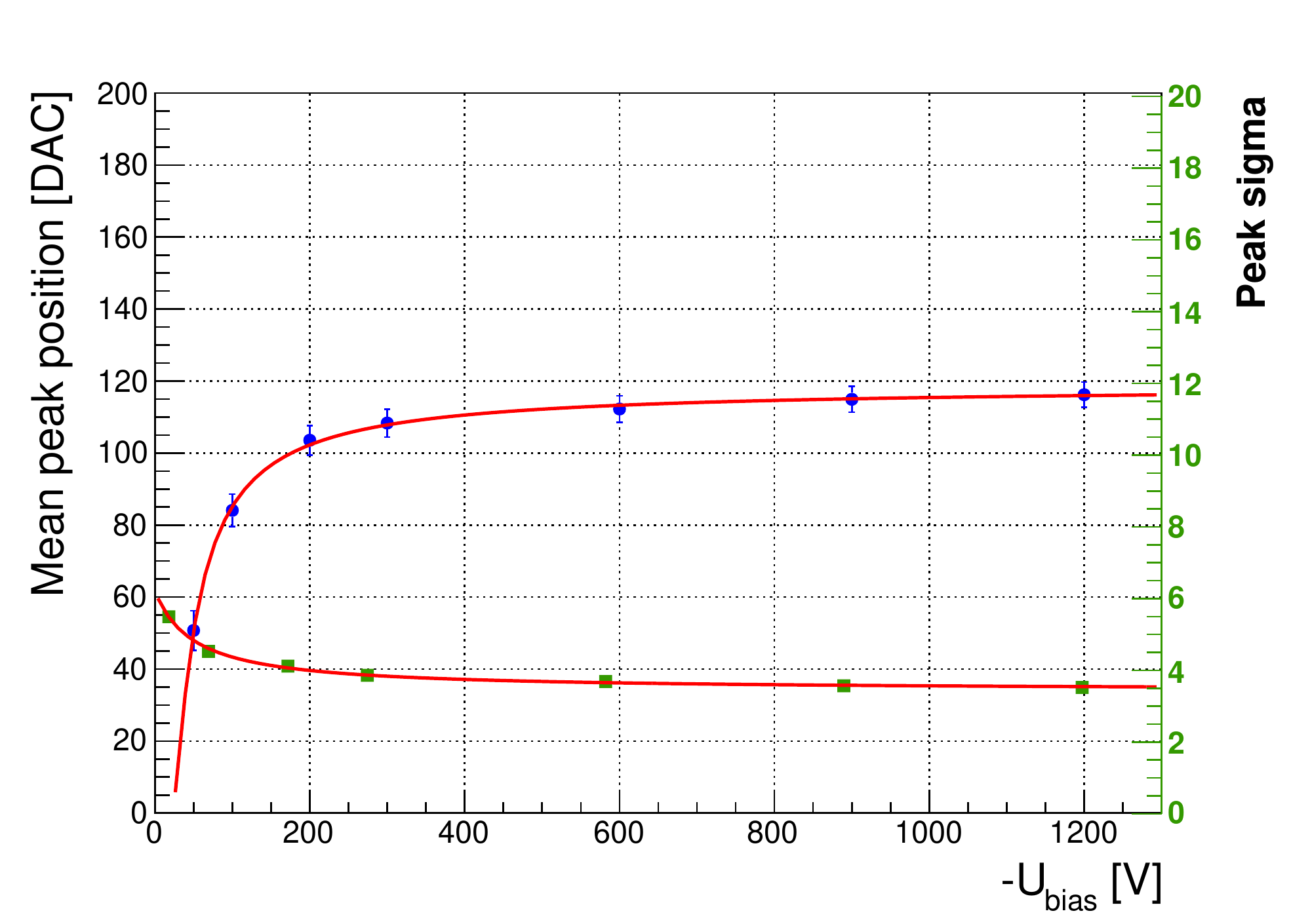}
\caption{\label{fig_PeakPos_biasdep} Dependencies of the indium photopeak mean value (blue markers) and the sigma of the peak (green markers) on the bias voltage measured by 500~$\mu m$ thick GaAs:Cr-based Timepix detector operating in TOT mode. Axis corresponding to the dependence of the dispersion is shown in green color.}
\end{figure}

The operating point of the bias voltage for 500~$\mu m$ thick GaAs:Cr-based Timepix detector in the Time-over-Threshold mode was set around $-300~V$. Indeed, as one can see in Figure~\ref{fig_PeakPos_biasdep}, the position of the indium photopeak at this bias voltage is already on the plateau and the sigma of the spectrum peak does not change significantly with higher bias voltage, while the leakage current is on the linear part of the IV curve.

But, when switching to the Medipix mode the bias voltage of -300~$V$ is not enough for good performance of the device. In order to determine the energy resolution of the GaAs-based Timepix detector operating in the Medipix mode the threshold scans were done for different values of the bias voltage up to $-1.2~kV$. Thus the influence of the bias voltage on the spectroscopic performance was investigated. The GaAs-based Timepix detector D08-W0153 with sensor thickness of 500~$\mu m$ was irradiated using fluorescence photons from an indium foil ($K_{\alpha_{1}}=24.2$ $keV$, $K_{\beta_{1}}=27.3$ $keV$). In order to get more accurate results by subtracting the background radiation in the X-ray safety cabinet, the threshold scans without foil were performed for each bias voltage setting. Also a data run with a low X-ray tube intensity has been taken for each bias voltage step.

\begin{figure}[htbp]
\centering 
\includegraphics[width=.45\textwidth]{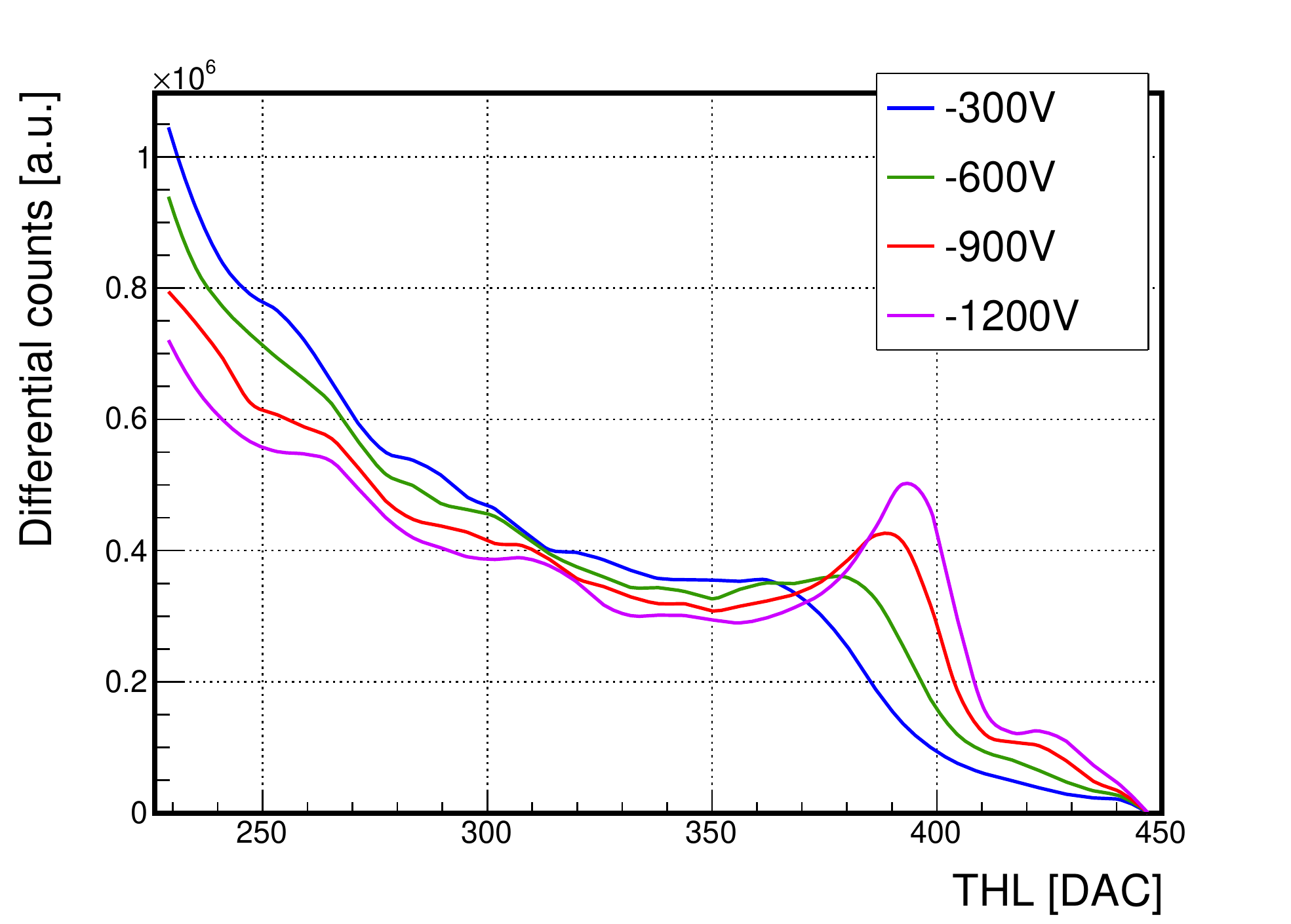}
\qquad
\includegraphics[width=.45\textwidth]{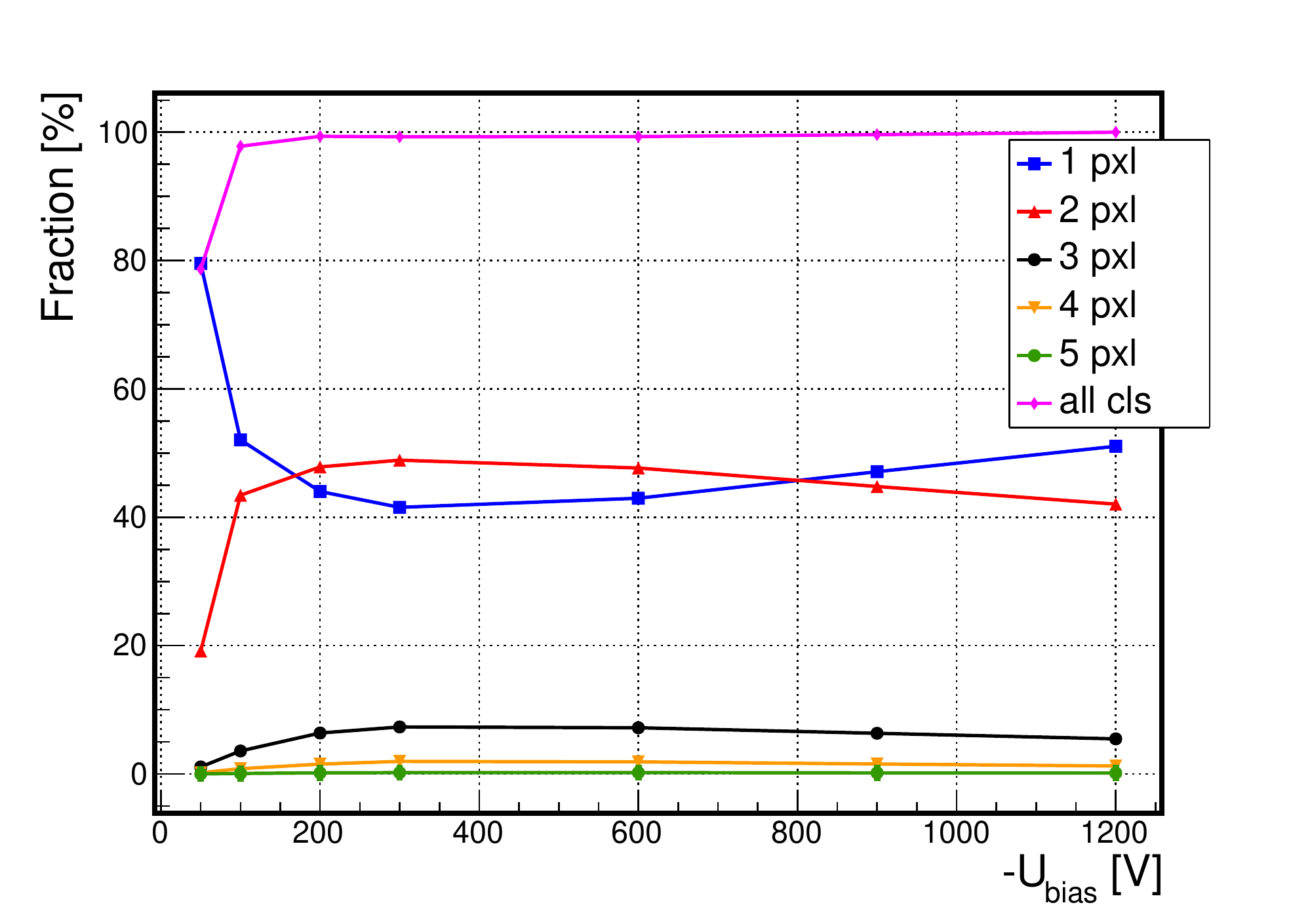}
\caption{\label{fig_ClsSize_ERF_bias} (left) Energy response function for fluorescence photons from an indium foil for different bias voltages measured by 500 $\mu m$ thick GaAs:Cr detector. (right) Percentage ratio of clusters of different sizes in dependence of the bias voltage.}
\end{figure}

The energy response function for fluorescence photons from indium foils for different bias voltages are shown in Figure~\ref{fig_ClsSize_ERF_bias} (left). The most significant thing here is the increase in the number of events in the photopeak with higher bias voltage. While at the voltage of $-300~V$ there is practically no visible peak, at $-1200~V$ almost gaussian shape peak is present. The photopeak corresponding to $K_{\beta}$-line of In and $K_{\alpha}$ fluorescence lines of Ga and As also become recognizable at $-1200~V$. It is connected with a smaller contribution of charge cloud diffusion during the drift processes at higher charge drift velocity. This explanation is confirmed by the dependencies shown in Figure~\ref{fig_ClsSize_ERF_bias} (right). Here the ratio of clusters with pixels from 1 to 5 is shown in dependence on the bias voltage. Firstly, number of single pixel events decreases with increasing the bias voltage which causes the growth of charge collection efficiency and transition of such events to double pixel clusters (a part of the shared charge induced on the second pixel exceeds the threshold and is registered by electronics). On the other hand, from $\sim~300~V$ the reverse process begins: number of single pixels events is increasing and double and even triple pixel events become less frequent with higher bias voltage. Such behavior has a positive impact on the detector performance: the multiple counting of single photons due to charge sharing is getting smaller.

\begin{figure}[htbp]
\centering 
\includegraphics[width=.8\textwidth]{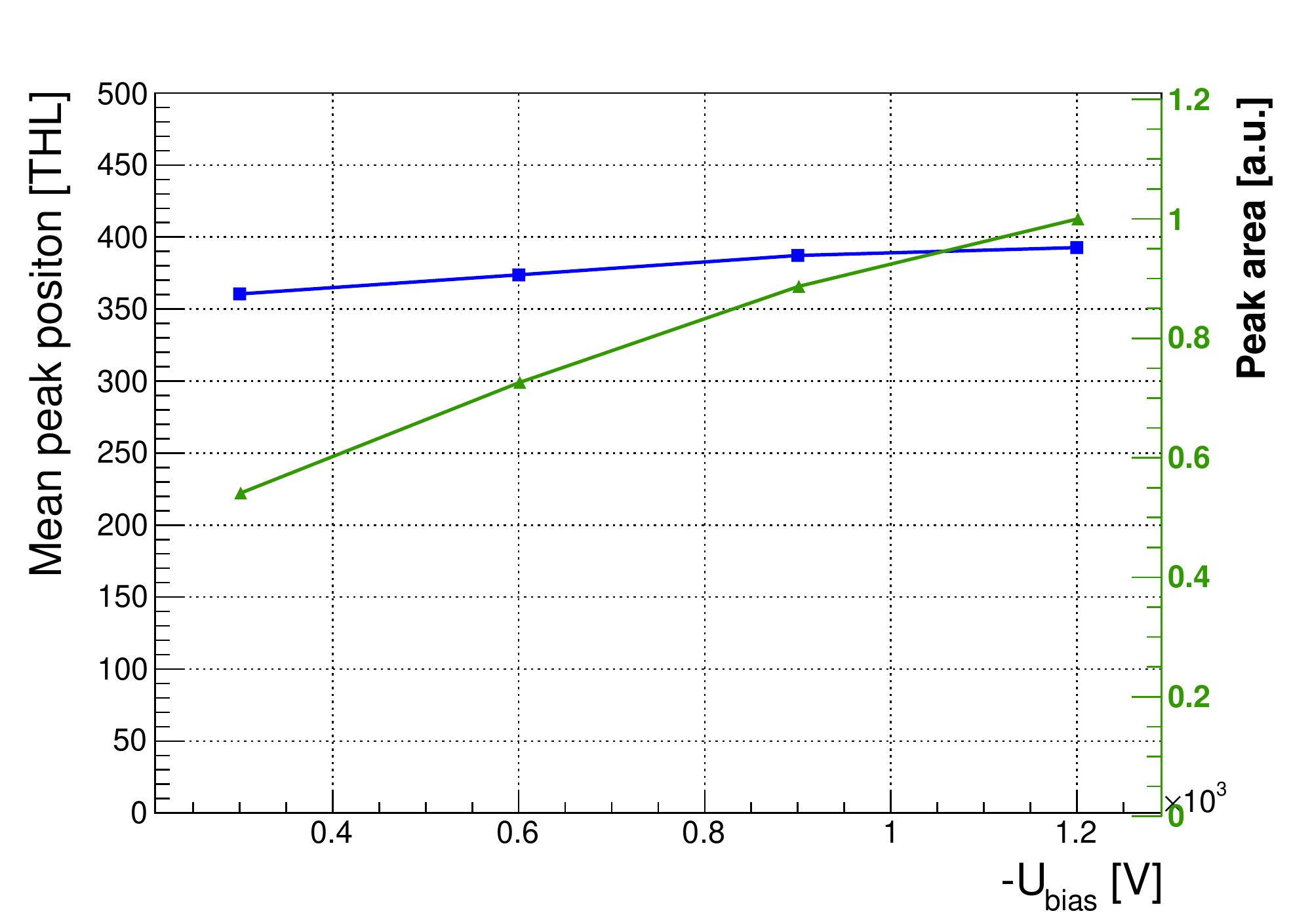}
\caption{\label{fig_PeakPos_biasdep_MPX} Dependencies of the indium photopeak mean value (blue markers) and normalized area under the peak (green markers) on the bias voltage measured by 500~$\mu m$ thick GaAs:Cr-based Timepix detector operating in the Medipix mode. Axis corresponding to the dependence of the peak area is shown in green color.}
\end{figure}

Since the energy peaks in the Medipix mode are not gaussian shaped (especially at low voltages), the area under photopeak is selected as criteria of the energy resolution. In Figure~\ref{fig_PeakPos_biasdep_MPX} these values are shown for different bias voltages together with the positions of photopeak. One can see that not only photopeak moves to higher energies with increasing bias voltage but also the number of events in this photopeak (twice more for $-1.2~kV$ relatively to $-300~V$). Presumably this trend continues at higher voltages, but the electrical field strength of $2.4~kV/cm$ is very close to the breakdown value in the air.   


\section{Spatial resolution for different thicknesses}

To determine the spatial resolution of the GaAs:Cr-based pixel detectors, measurements of the presampling modular transfer function (MTF) were carried out using the slanted edge method~\cite{samei1998method}. In order to study the influence of the detector thickness on the spatial resolution the MTF was measured for 300, 500, 900 $\mu m$ thick detectors. The sharp edge of a 200~$\mu m$ thick tungsten mask and X-ray tube with a 15 $\mu m$ focal spot operating at 30 $kV$ have been used\footnote{Hamamatsu L10951}. The mask was placed at an angle of $3^o$ with respect to the pixel columns of the Timepix detector. To avoid blurring due to geometrical unsharpness the mask was placed on the top of the sensor. 

\begin{figure}[htbp]
\centering 
\includegraphics[width=.9\textwidth]{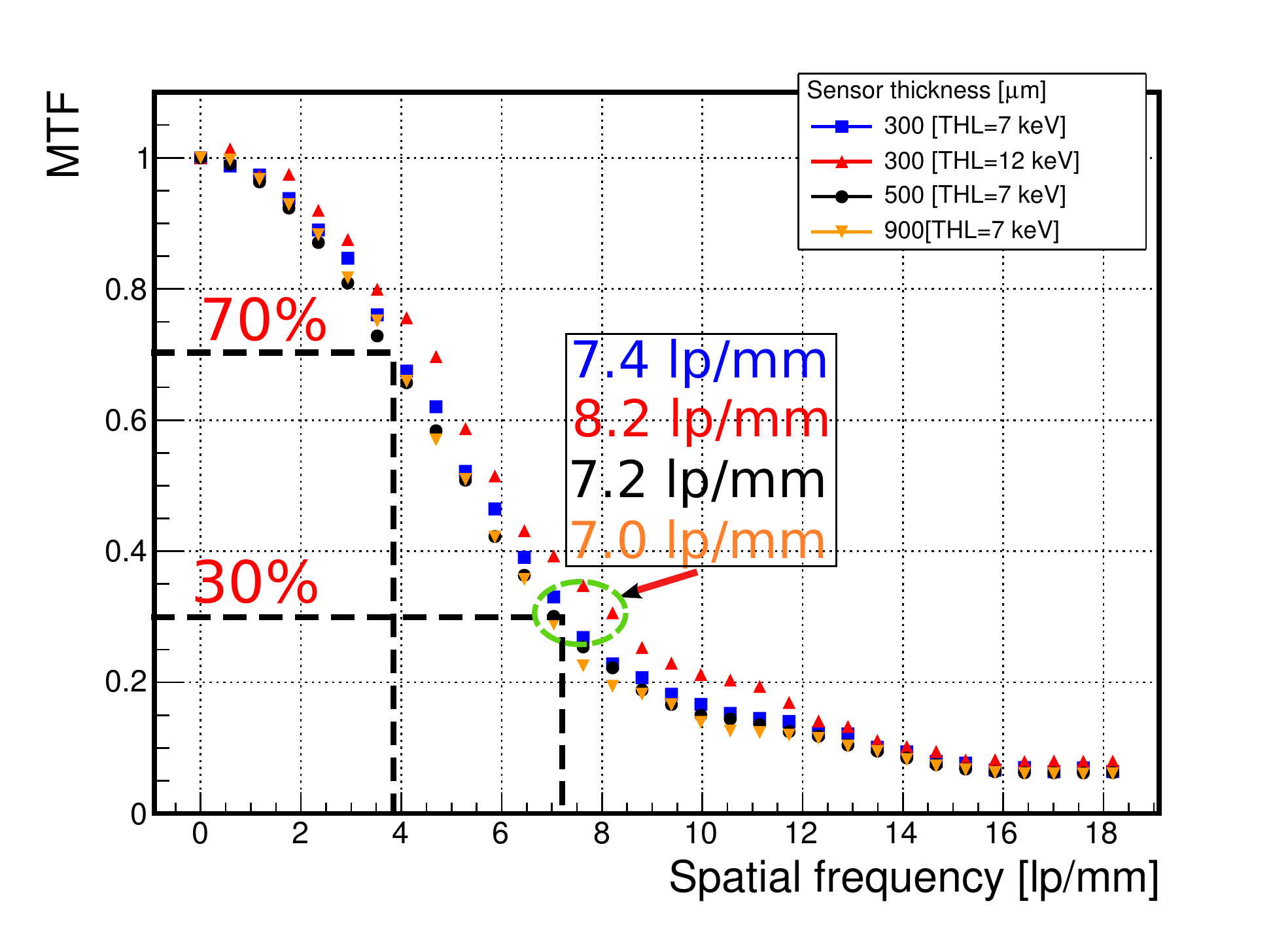}
\caption{\label{fig_SpatialRes} Presampling of MTF for 300, 500 and 900 $\mu m$ thick sensors and for two threshold levels: 7 $keV$ and 12 $keV$.}
\end{figure}

In order to study the influence of charge sharing on the spatial resolution the MTF was calculated\footnote{The ImageJ plugin "SE MTF"~\cite{imagej} was used for MTF calculations} for detectors of different thickness and for two threshold levels. Figure~\ref{fig_SpatialRes} shows the MTF calculation results. The level selected as characteristic of the spatial resolution is the spatial frequency at which the MTF decreases to 30~\% of its maximum. The influence of the threshold value on the MTF is clearly visible for a 300~$\mu m$ thick detector: at the same photon energy the MTF for the lower threshold level (7 $keV$) gives a worse spatial resolution of 7.4 $lp/mm$ than the higher threshold level (12 $keV$) -- 8.2 $lp/mm$. It is a consequence of charge sharing and presence of fluorescence photons from GaAs:Cr. The worst spatial resolution is obtained for the Timepix detector with a 900~$\mu m$ thick GaAs:Cr sensor (7.0 $lp/mm$), which is also connected with a higher probability of fluorescence photons emission. 



\section{Conclusions}

In this work the results of investigations of the Timepix detector with the GaAs:Cr sensor have been presented. The main characteristics of the novel detectors were systematically evaluated over the last five years: the volt-ampere characteristics for various temperatures; detector homogeneity and stability under action of photon fluxes; the energy and spatial resolution of detectors with various sensor thicknesses operating in the Medipix and Time-over-Threshold modes; temperature dependence of detector characteristics. It was shown that by optimizing the bias voltage it is possible to achieve in the Medipix mode a FWHM of 2.5 $keV$ at 25 $keV$ for 500 $\mu m$ thick detector despite of the charge sharing effect. Measured map of electron $\mu \tau$-product demonstrates an acceptable uniformity of the transport characteristics over the sensor matrix.  

We conclude that properties of GaAs:Cr are suitable for its use in X-ray imaging detectors along with the well-known Si, thus extending the photon absorption efficiency range up to 60 $keV$. But, applications of thick ($>$ 500 $\mu m$) GaAs:Cr sensors for spectroscopic X-ray imaging are limited by a strong influence of the charge sharing effect, which can be compensated by offline data processing (the high speed readout is necessity in this case) or by using novel readout chips which offer inter-pixel communication (e.g. the Medipix3RX~chip~\cite{ballabriga2013medipix3rx}).

\section{Acknowledgments}
This work was supported by the Ministry of Education and Science of the Russian Federation under Contract No. 14.618.21.0001 in the framework of the project RFMEFI61814X0001. The authors would like to express their gratitude to colleagues from the Department of Colliding Beam Physics of the JINR LNP and from the Institute of Experimental and Applied Physics in Prague for remarks and suggestions on this work and for the support during the measurements. Also we would like to thank the members of the Medipix team for very useful discussions.  




\end{document}